\documentclass[twocolumn,notitlepage,showpacs,pre,preprintnumbers,amsmath,amssymb,aps,floatfix]{revtex4-1}
\usepackage{graphicx}
\usepackage{dcolumn}
\usepackage{bm}
\usepackage{soul}
\usepackage{color}

\newcommand{\eps}{\varepsilon}

\newcommand{\BEQ}{\begin{equation}}
\newcommand{\EEQ}{\end{equation}}
\newcommand{\BEA}{\begin{eqnarray}}
\newcommand{\EEA}{\end{eqnarray}}

\newcommand{\nn}{\nonumber}

\begin{document}
\title{Noise-induced transitions in gene circuits: a perturbative approach for slow noise}
\author{Gerardo Aquino$^{1}$}
\email{Gerardo.Aquino@apha.gov.uk}
\author{Andrea Rocco$^{2,3}$}
\email{A.Rocco@surrey.ac.uk}
\affiliation{$^{1}$Department of Computing,  Goldsmiths, University of London, London, UK}
\affiliation{$^{2}$Department of Physics, University of Surrey, GU2 7XH Guildford, UK}
\affiliation{$^{3}$Department of Microbial Sciences, University of Surrey, GU2 7XH, Guildford, UK}

\begin{abstract}
We consider a generic class of gene circuits affected by nonlinear extrinsic noise. To address this nonlinearity we introduce a general perturbative methodology based on assuming timescale separation between noise and genes dynamics, with fluctuations exhibiting a large but finite correlation time. We apply this methodology to the case of the toggle switch, and by considering biologically relevant log-normal fluctuations, we find that the system exhibits noise-induced transitions. The system becomes bimodal in regions of the parameter space where it would be deterministically monostable. We show that by including higher order corrections our methodology allows one to obtain correct predictions for the occurrence of transitions even for not so large correlation time of the fluctuations, overcoming thereby limitations of previous theoretical approaches. Interestingly  we find that at intermediate noise intensities the noise-induced transition in the toggle switch affects one of the genes involved, but not the other one. \end{abstract}

\maketitle

\section{Introduction}

Extrinsic noise has been investigated thoroughly over the last decades in a variety of physical and chemical systems \cite{Horsthemke84}. It has emerged as an active dynamical player, which can produce so-called noise-induced transitions to bimodal dynamics in systems otherwise deterministically monostable.

In chemical and physical systems, the search and the possible emergence of such transitions relies on the multiplicative nature of the corresponding stochastic dynamics, which is mathematically well characterized when the noise appears linearly in the system’s dynamical equations, and   it is gaussian and white. In this case the Stratonovich prescription to interpret the corresponding stochastic integral leads to the emergence of the so-called Stratonovich drift in the Langevin and Fokker-Planck dynamics, which is responsible for dramatic changes in the stability properties of stationary states, and possibly in their number \cite{Gardiner85}.

In biological systems, however, none of the assumptions on the noise being linear, Gaussian and white is usually fulfilled. In gene regulatory networks, for instance, the noise is often non-Gaussian \cite{Cai06,Bengtsson05}, it is usually characterized by correlation times that can easily exceed cell cycles times \cite{Rosenfeld05}, and it appears mostly in a nonlinear fashion in the dynamics determining gene expression.
 
While it can be very hard to treat exactly the corresponding stochastic dynamics under these conditions, a geometric construction based on noise filtering has been  introduced \cite{Ochab-Marcinek10,Ochab-Marcinek17}. In essence, this approach relies on the transformation properties of probability distributions of  stochastic variables related by a given input/output transformation. Conservation of probability allows one to express the output probability distribution in terms of the input probability distribution, which results in non-trivial dynamics when the relation between the stochastic variables is nonlinear.

This  {approach} has been rederived dynamically in \cite{Aquino20} by extending to nonlinear noise the so-called switching-curve approximation introduced in \cite{Arnold78,Reichl82}. The dynamical derivation highlights the crucial role played by the timescales of the system. The approximation works well when the input noise is much slower (in the limit, infinitely slower) than the dynamics of the system of interest. In a gene expression process, this translates in extrinsic noise being much slower that protein degradation and synthesis \cite{Aquino20}. 

In this paper we take up this issue again, and develop a perturbative framework that allows one to go beyond the approximation defined in \cite{Aquino20} by extending that approach further to large but finite correlation times of the noise. We show that the inclusion of higher order perturbative terms can modify dramatically the analysis of the emergence of noise-induced transitions, by removing them when present at lower orders, and providing therefore a qualitative and quantitative more accurate description of the stochastic dynamics.

We apply our formalism to the case of the toggle switch \cite{Gardner00}, and show that when log-normal noise is considered the system undergoes a transition in a region of the phase space where the deterministic system would be monostable. We also find that tuning the noise intensity may give rise to a situation in which the marginalized probabilities of the genes involved show different qualitative features, with overlapping unimodal and bimodal behaviours. We name this feature 'partial bimodality', and discuss its origin in the context of noise propagation across the network.

The paper is organized as follows. In Section II we present the genetic circuits considered here. In Section III we introduce our perturbative approach, and develop it up to the second perturbative order for the generic circuit introduced in Section II (with the third perturbative order described for the toggle switch in the Appendix). In Section IV we introduce the toggle switch as an application of the methodology, and assess the perturbative framework by comparing the contributions of different orders to direct stochastic simulations. In Section V we draw conclusions on the methodology introduced, and make some final remarks.

\section{Genetic circuits}
We consider a generic circuit of interacting genes. We make the assumption that the half-life of mRNAs is much shorter than that of the corresponding proteins, so that mRNA dynamics can be adiabatically eliminated, and gene expression can be described in a single step of combined transcription and translation \cite{Rosenfeld02}. We then describe the system by decomposing the rate equations for each synthesized protein into a production term and a degradation term as follows: 
\BEQ
\label{Gtogsw}
\frac{d  {\bf x}}{dt} ={\bf g}({\bf x},R) -  {\bf K} {\bf x} \ .
 \EEQ
Here ${\bf x}$ is an $n$-dimensional vector whose components are the $x_i$ species involved, while ${\bf K}$ is a diagonal matrix whose diagonal elements are the degradation rates $k_i=1/\tau_{x_i}$ of each species $x_i$, with $\tau_{x_i}$ the respective half-lifes. The function ${\bf g}({\bf x},R)$ is a $\mathbb{R}^n \times \mathbb{R}\to \mathbb{R}^n$ Hill-like function of the concentrations of the $n$ species ${\bf x}$, and $R$ is a factor that exerts control on the dynamics of ${\bf x}$. The factor $R$ can be thought of as an external control parameter (for instance a binding constant), or any other species at high enough concentrations so that it is unaffected by the dynamics of ${\bf x}$ (for instance an external signalling molecule, or a transcription factor). For the sake of simplicity, we here limit ourselves to a single factor $R$, but the procedure can be further generalised to multiple factors.
 
\section{Perturbative calculation}
Let us now consider fluctuations acting on $R$. We describe them by adding a generic Langevin-type dynamics to equations (\ref{Gtogsw}),
 \BEA
 \label{genneta}
\frac{d {\bf x}}{dt} &=&  {\bf g}({\bf x},R)-{\bf K}{\bf  x} \ ,\\
 \label{gennetb} \frac{d R}{dt}&=&\frac{\mu(R)}{\tau}+\sqrt{ \frac{D}{\tau} } \nu(R) \xi(t) \ , 
\EEA
with $\tau$ the timescale of the process $R$. In Eq. (\ref{gennetb}) the functions $\mu(R)$ and $\nu(R)$ are kept generic for now.  Their specification allows us to reproduce different noise distributions. Also, the noise $\xi(t)$ is assumed Gaussian and white, with correlator $\langle \xi(t) \xi(t^\prime) \rangle = 2 \delta (t - t^{\prime})$.

By rescaling time as $t \to t/\tau_M$, with  { $\tau_M=1/k_M$ } the largest degradation time of the species involved (we assume  that $\tau_{x_i} \lessapprox \tau_M$ for $x_i\neq x_M$) we obtain
\BEA
\label{noise2a} \frac{d {\bf x}}{dt} &=& \eps {\boldsymbol{ \gamma}}({\bf x},R)-{\bf \bar{K}}{\bf  x} \ ,\\
 \label{noise2b} \frac{d R}{dt}&=&\eps\mu(R)+\sqrt{ \eps D } \nu(R) \xi(t) \ ,
\EEA
where $\eps=\tau_M/\tau$,    {${\boldsymbol{ \gamma}}= \tau {\bf g} $} and ${\bf \bar{K}}={\bf K}/k_M$ with ${\bf \bar{K}}$ still diagonal and
${\bf \bar{K}}_{ii}=\kappa_{ii}=k_{x_i}/k_M=O(1)$.  {In order for the dynamics of $R$ to be slower than that of all the species $x_i$, we consider the condition $\tau_M \ll \tau$, implying that  $\eps$ is a small parameter, with $\eps \ll 1$.}

 {This naturally leads to seeking a solution of the system
  (\ref{noise2a}) and (\ref{noise2b}) in terms of a perturbative expansion in orders of $\eps$. This perturbative approach is inspired by the one presented in \cite{Aquino20}, with the important difference that the latter was based on assuming the limit $\tau \to \infty$, while }  {here we consider instead the limit  $\tau_M \to 0$ (i.e. $ \eps \to 0$) with $\tau$ large but fixed}  {and finite}.  {This different perspective}  {keeps the function ${\boldsymbol{ \gamma}}= \tau {\bf g} $ in (\ref{noise2a}) fixed in the limit $ \eps \to 0$},  {and allows us to obtain a systematic perturbative expansion in $\eps$. As a result, the approach presented here allows us to go beyond the lowest order calculation of \cite{Aquino20}, accounting for the noise filtering approach of \cite{Ochab-Marcinek10,Ochab-Marcinek17}, and to obtain all perturbative corrections analytically at any order in $\eps$.}

Let us define the Fokker-Planck operator $L_{\rm FP}$ for the $R$ process as
\BEQ
\label{FP}
L_{\rm FP}(R)=\frac{\partial}{\partial R}\left[- (\mu+ D \nu\nu')+  D\frac{\partial}{\partial R}\left( \nu^2 \right) \right] \ , 
\EEQ
where we have assumed the Stratonovich interpretation for the multiplicative noise equation (\ref{noise2b}).

The total Fokker-Planck equation for the time-dependent probability density $W_t({\bf x},R)$ is therefore:    
\begin{eqnarray}
\label{totalb0}
\nonumber \frac{\partial}{\partial t} W_t({\bf x},R)&=&\left[-\mathbf{\nabla} \cdot  \left(  \eps{\boldsymbol{ \gamma}}({\bf x},R)-{\bf \bar{K}} {\bf  x}\right) \right. \\
&& \qquad \quad \left.   +\eps L_{\rm FP}(R) \right] W_t({\bf x},R) \ .
\end{eqnarray}
We now aim to compute the stationary solution of Eq. (\ref{totalb0}), namely the time-independent probability density $W({\bf x},R)$ satisfying 
\BEA
\label{sstateZ}
\left[\mathbf{\nabla} \cdot  \left(  \eps{\boldsymbol{ \gamma}}({\bf x},R)-{\bf \bar{K}} {\bf  x}\right) 
  -\eps L_{\rm FP}(R) \right] W({\bf x},R)
=0 \ ,
\EEA
obtained by imposing $\frac{\partial}{\partial t} W_t({\bf x},R)=0$ in (\ref{totalb0}).
We seek a solution of the type
\BEQ
\label{FUN}
W({\bf x},R)=\bar{W}(R)\psi({\bf x},R)
\EEQ
where $\bar{W}$ is chosen so that $L_{\rm FP} \bar{W}(R)=0$, since the dynamics of the variable $R$ is independent of $x,y$ (but not viceversa). In general it can be shown \cite{Gardiner85} that $\bar{W}(R)$ can be written as 
 \BEQ
 \label{generalpf}
 \bar{W}(R)=\frac{\mathcal{N}}{\nu(R)}\exp \left(\frac{1}{D}\int^R dR'\frac{\mu(R')}{\nu^2(R')}\right) \ ,    
\EEQ
and it must  be as well
 \BEQ
\label{FUNINT}
\int {\bf dx}W({\bf x},R)= \bar{W}(R)  \Rightarrow  \int {\bf dx}  \psi({\bf x},R)= 1 \ .
\EEQ 

\subsection{Zeroth order solution}
We begin by assuming the following perturbative expansion for $W({\bf x},R)$ of generic order $m$ in powers of $\eps$ 
  \BEQ
     \label{expansionZ}
     W({\bf x},R)=\sum^m_{j=0} \frac{\eps^j}{j!} W^{(j)} ({\bf x},R)+O(\eps^{m+1}) \ .
\EEQ
By replacing this expansion in the lefthand side of (\ref{sstateZ}), and by keeping  only the zeroth order terms in $\eps$ we obtain
\BEQ
\left( \mathbf{\nabla} \cdot {\bf x} \right)W^{(0)}({\bf x},R) =0 \ , \label{nabla}
\EEQ
which immediately leads to the zeroth order solution
\BEQ
\label{W0z}
W^{(0)} ({\bf x},R)=\delta^n({\bf x})\bar{W}(R)=\bar{W}(R) \prod_{i=1}^n\delta(x_i) \ .
\EEQ

\subsection{First order correction}
Having evaluated $W^{(0)}$ we can replace again the expansion (\ref{expansionZ}) in the lefthand side of (\ref{sstateZ}), and keep this time all terms up to first order in $\eps$. We obtain:
\BEA
\nonumber &&- \mathbf{\nabla} \cdot  \left( {\boldsymbol{ \gamma}}({\bf x},R)W^{(0)} - {\bf \bar{K}}{\bf x} W^{(1)}\right)+L_{\rm FP}(R) W^{(0)}=0 \ ,
\EEA
which, using the property $x  \frac{d^n \delta(x)}{dx}=-n \frac{d^{n-1} \delta(x)}{dx}$ and noticing that $L_{\rm FP}(R) W^{(0)}=0$, leads to the solution
\BEQ
\label{W1z}
W^{(1)} ({\bf x},R)=-\bar{W}(R) \sum_{i} \left( \frac{\delta'(x_i) }{\kappa_{ii}}\gamma_i({\bf x},R) \prod_{j\neq i} \delta(x_j) \right) \ .
\EEQ
Therefore, up to the first  order in $\eps$ we can write 
\BEA
\label{firstSolz}
&& W({\bf x},R) \nonumber \\ 
&&\quad =W^{(0)}({\bf x},R)+\eps W^{(1)}({\bf x},R)\nonumber \\
&&\quad = \bar{W}(R) \left[ \vphantom{\frac{\delta'(x_i) }{\kappa_{ii}}}\delta^n({\bf x}) - \eps \sum_{i} \left( \frac{\delta'(x_i) }{\kappa_{ii}}\gamma_i({\bf x},R) \prod_{j\neq i} \delta(x_j) \right)\right] \nonumber \\
&&\quad \simeq \bar{W}(R) \; \delta^n \left({\bf x}- {\frac{\eps}{\bf \bar{K}}}  {\boldsymbol{ \gamma}}({\bf x},R) \right) \ ,
\EEA
where the last equality is meant in the sense of distributions, and is also valid up to the first order in $\eps$:
\BEA
&&\hspace{-0.6cm}\int d{\bf x} \delta^n\left({\bf x}- \eps\frac{1}{\bf \bar{K}} {\boldsymbol{ \gamma}}({\bf x},R) \right) h({\bf x}) =\int d{\bf x} \left[ \vphantom{\frac{\delta'(x_i) }{\kappa_{ii}}} \delta^n({\bf x}) \right. \nonumber \\
&&\hspace{-0.4cm} - \left.  \eps \sum_i \left( \frac{\delta'(x_i) }{\kappa_{ii}}\gamma_i({\bf x},R) \prod_{j\neq i} \delta(x_j) \right)  \right]  h({\bf x}) +O(\eps^2) \,
  \EEA
for any test function $h({\bf x})$. We notice here that this first order result coincides with the   result of the nonlinear noise filtering approach proposed in \cite{Ochab-Marcinek10} and derived dynamically in \cite{Aquino20}.
In fact the delta function in Eq. (\ref{firstSolz})  defines the (approximate) steady state solution, determined by the condition
 \BEQ
\label{oneimplicit}
{\bf z}({\bf x},R)={\bf x}-\eps{\frac{1}{\bf \bar{K}}}  {\boldsymbol{ \gamma}}({\bf x},R)=0.
\EEQ
This solution coincides with the  steady state solution of Eqs. (\ref{genneta}) and (\ref{gennetb}) in the limit $\tau \to \infty$  of  infinitely slow noise (i.e. constant $R$ on the time scale of  the dynamics of  ${\bf x}$)
and  corresponds trivially to a  change of variable $R \to {\bf x}$.

For instance, in the one-dimensional case, when $\gamma(x,R)=\gamma(R)$ (i.e. $g(x,R)=g(R)) $ depends monotonically only on $R$,  
 one can use  the one-dimensional version of  Eq. (\ref{firstSolz}) and after integrating over $R$ via
\BEA
\nonumber p(x)&=&\int dR W(x,R)=\int dR \bar{W}(R) \delta(x-\frac{\eps}{\kappa} \gamma(x,R))\\
&&=\int dR \bar{W}(R) \delta(x-\frac{\eps}{\kappa}\gamma(R)),  
\EEA
 obtain the transformation of the probability density function
\BEQ
\label{1dim}
p({ x})=\bar{W}(r({ x})) |dr({ x})/d{ x}|.   
\EEQ
In Eq. (\ref{1dim})  $r({x})$ descends from the one-dimensional version of (\ref{oneimplicit}), i.e. $x=\frac{\eps}{\kappa} \gamma(R)$, and therefore $r(x)=R=\gamma^{-1}(\kappa x/\eps)={g}^{-1}(k x)$, with $\kappa=k/k_{M}$ being the only value of the one-dimensional matrix $ {\bf \bar{K}}$.

This solution corresponds to the nonlinear filtering approach introduced in \cite{Ochab-Marcinek10,Ochab-Marcinek17}, where the Hill-like $g$ function acts as transfer function of the regulator $R$,  assumed as an infinitely slow fluctuating variable, and produces a corresponding distribution of stationary protein concentration ${x}$. 

However, this perturbative approach can be extended to describe noise with finite  $\tau$ beyond the first order, providing non-trivial corrections to the resulting dynamics as it is shown in the next subsection, as well as to generic $n$-dimensional systems by using the implicit functions theorem.

 
In the multi-dimensional case,  in fact, when ${\bf z}({\bf x},R)$ is a continuously differentiable function $ \mathbb{R}^{n+1} \to \mathbb{R}^n$, the implicit functions theorem tells us that a  function $r({\bf x})=R$  exists which is  locally invertible so that ${\bf z}({ r^{-1}}(R),R)$=$0$.
It follows that the probability density for ${\bf x}$ can be determined from $ \bar{W}(R) $ via a change of variable as
\BEQ
\label{ndimgen}
p({\bf x})= \bar{W}(r({\bf x}) )|J |.
\EEQ
Here $|J|$ is the Jacobian determinant  of the transformation $(R=r({\bf x}), x_2 \cdots x_n)\to(x_1 \cdots x_n) $ 
which can be obtained, without  knowing the explicit form of $r$,  from the function ${\bf z}$ in Eq. Eq. (\ref{oneimplicit}) using the implicit function theorem  in $n$ dimensions.  {Eq. (\ref{ndimgen}) generalises Eq. (\ref{1dim}) to the case of $n$ species, with $r({\bf x})$ now depending on all species  $x_1 \cdots x_n$ and defined only implicitly via Eq. (\ref{oneimplicit}).}
 

\subsection{Second order correction}
In order to evaluate the second order correction we need first to evaluate  $L_{\rm FP} (R) W^{(1)} ({\bf x},R)$.
From Eq. (\ref{W1z})  and the property $L_{\rm FP} (R) \bar{W}(R)=0$
it follows that 
\BEA
\label{L1z}
&&-L_{\rm FP}(R) W^{(1)}({\bf x},R) \nonumber \\
&&\qquad =\sum_i \frac{\delta'(x_i)}{\kappa_{ii}}\prod_{k\neq i}
\delta(x_k) L_{\rm FP}(R) \left[\gamma_i({\bf x},R) \bar{W}(R)\right] \nonumber \\
&&\qquad =\sum_i \frac{\delta'(x_i)}{\kappa_{ii}}G_i({\bf x},R)\prod_{k\neq i}\delta(x_k)  
\EEA
where $G_i({\bf x},R)=L_{\rm FP}(R) \left[\gamma_i({\bf x},R) \bar{W}(R)\right]$ which, using Eq.(\ref{FP}), leads to:
\BEA
\nonumber &&\hspace{-0.6cm}G_i({\bf x},R) \nonumber \\
&& \hspace{-0.6cm} = - (\mu(R)-3 D\nu(R) \nu'(R))\bar{W}(R)\partial_R \gamma_i({\bf x},R) \nonumber \\
&& \hspace{-0.6cm} \quad  + D \nu(R)^2 (2\bar{W}'(R)  \partial_R \gamma_i({\bf x},R) +\bar{W}(R) \partial_R^2 \gamma_i({\bf x},R) \ .
\EEA

Having evaluated $W^{(1)} ({\bf x},R)$ we can now repeat the same procedure as done for the first order, {\em i.e.} replace expansion (\ref{expansionZ})  in the lefthand side of (\ref{sstateZ}) and keep this time all terms up to second order in $\eps$. We obtain:
\BEQ
- \mathbf{\nabla} \!\cdot\!  \left( {\boldsymbol{ \gamma}}({\bf x},R)W^{(1)} - {\bf \bar{K}}{\bf x} \frac{W^{(2)}}{2}\right)+L_{\rm FP}(R) W^{(1)} ({\bf x},R)=0 \ , \label{new}
\EEQ
which, using (\ref{L1z}), can be rewritten as
\BEA
\label{secordz}
&&\mathbf{\nabla} \cdot  \left( {\boldsymbol{ \gamma}}({\bf x},R)W^{(1)} - {\bf \bar{K}}{\bf x} \frac{W^{(2)}}{2}\right) \nonumber \\
&& \qquad\qquad +\sum_i \frac{\delta'(x_i)}{k_{ii}} G_i({\bf x},R)\prod_{k\neq i}\delta(x_k) =0 \ .
\EEA
From the structure of this equation and from (\ref{firstSolz}) we deduce that the solution for $W^{(2)}({\bf x},R)$ must have  the following form:
\BEA
\label{secordbz}
W^{(2)}({\bf x},R)&=&\bar{W}(R)\sum_{i,q} \left[\left(A_{iq} \delta'(x_i)+B_{iq}\delta^{\prime \prime}(x_i)\right)\delta(x_q) \right. \nonumber \\
&&\left. \qquad + C_{iq}\delta'(x_i)\delta'(x_q)\right] \prod_{k\neq i,q}\delta(x_k) \ .
\EEA
The coefficients $A_{iq}$, $B_{iq}$, $C_{iq}$ can be obtained by inserting (\ref{secordbz}) in (\ref{secordz}) and result in
\begin{subequations}
\label{secordcz2}
 \begin{align}
 A_{iq}&=\left(-\frac{2}{k_{ii}^2}\bar{G}_{i}({\bf x},R)+\tilde{A}_{i}({\bf x},R)\right) \delta_{iq}  \label{secordcz2a}\\
\nonumber & \text{with} \;\;\frac{\tilde{A}_{i} ({\bf x},R)}{x_i}-\partial_{x_i} \tilde{A}_{i}({\bf x},R)=-\frac{2}{\kappa_{ii}^2}\partial_{x_i}\bar{G}_{i}({\bf x},R)\\
 B_{iq}&=\left(\frac{\gamma_i({\bf x},R)}{\kappa_{ii}}\right)^2 \delta_{iq}  \label{secordcz2b} \\
C_{iq}&=\frac{\gamma_i({\bf x},R)\gamma_q({\bf x},R)}{\kappa_{qq} \kappa_{ii}} (1-\delta_{iq}) \label{secordcz2c}
\end{align}
\end{subequations}
with $\bar{G}_i=G_i/\bar{W}(R)$ .
Therefore we make the following {\em ansatz} for the stationary probability density $W({\bf x},R)$:
\BEQ
\label{ansatzz2}
 W({\bf x},R)=\bar{W}(R) \delta^n\left({\bf x}-\frac{\eps}{\bf \bar{K}} {\boldsymbol{ \gamma}}({\bf x},R)-\frac{\eps^2}{2}{\bf Q}({\bf x},R)\right)  {+O(\eps^3)}
\EEQ
with
\BEQ
{\bf Q }({\bf x},R)=\left(- \frac{2 }{ \bf \bar{K}^2 }{\bf \bar{G}}({\bf x},R)+{\bf \tilde{A}}({\bf x},R)\right).
\EEQ
The solution given by the ansatz (\ref{ansatzz2}) in fact coincides up to second order in $\eps$ with the solution derived with the  perturbative expansion (\ref{expansionZ})  to the same order, i.e. including  Eq. (\ref{secordbz}) with coefficients (\ref{secordcz2}).

The delta function in Eq. (\ref{ansatzz2})  {is similar to the one in (\ref{firstSolz}) and therefore},  {as it was seen} for the first order in (\ref{oneimplicit}), it defines the (approximate) steady state solution  {of Eqs. (\ref{genneta}) and (\ref{gennetb})}.  {This is} determined by the condition
\BEQ
\label{dinizz}
{\bf z}({\bf x},R)={\bf x}-\frac{\eps}{\bf \bar{K}} {\boldsymbol{ \gamma}}({\bf x},R)-\frac{\eps^2}{2} {\bf Q}({\bf x},R)=0,
\EEQ
which in the case that $ \mathbf{\nabla} \cdot \frac{1}{\bf \bar{K}}{\boldsymbol{ \gamma}}({\bf x},R)=const$, implying in turn $ \mathbf{\nabla} \cdot\frac{1}{\bf \bar{K}} {\bf G}({\bf x},R)=0$ and therefore (neglecting null contributions in distributional sense) ${\bf \tilde{A}}({\bf x},R)=0$, simplifies to
\BEQ
\label{diniz}
{\bf z}({\bf x},R)={\bf x}-\frac{\eps}{\bf \bar{K}} {\boldsymbol{ \gamma}}({\bf x},R)-\eps^2 \frac{1 }{ \bf \bar{K}^2 }{\bf \bar{G}}({\bf x},R)=0.
\EEQ

{Note that both   ${\bf Q}({\bf x},R)$  and ${\bf \bar{G}}({\bf x},R)$ depend linearly on  ${\boldsymbol{ \gamma}}=\tau {\bf g}$. Therefore, when multiplied by their expansion factor $\eps^2=\frac{\tau_M^2}{\tau^2}$, they disappear in the limit $\tau \to \infty$, leaving only the first order correction in  Eqs. (\ref{ansatzz2}), (\ref{dinizz}) or (\ref{diniz}).  
This can be verified to apply to all the higher order corrections in the same limit. Therefore  Eq. (\ref{oneimplicit}), obtained as first order correction within our approach,  is actually  an exact solution in such limit,}  {corresponding thereby to the noise filtering approach of \cite{Ochab-Marcinek10,Ochab-Marcinek17}.}

Under the assumption that ${\bf z}({\bf x},R)$ is a continuously differentiable function $ \mathbb{R}^{n+1} \to \mathbb{R}^n$, the implicit functions theorem can be applied again. As a result, the probability density for ${\bf x}$ can be determined from $ \bar{W}(R) $ via a change of variable  as shown in Eq. (\ref{ndimgen}),
where the Jacobian can now be obtained from the function ${\bf z}$ in Eq. (\ref{dinizz}) or  (\ref{diniz}). 

In the following section, we show how this approach is applied with a concrete example.

\section{The toggle switch}

 \begin{figure}[t!]
\includegraphics[height=3 cm,  angle=0]{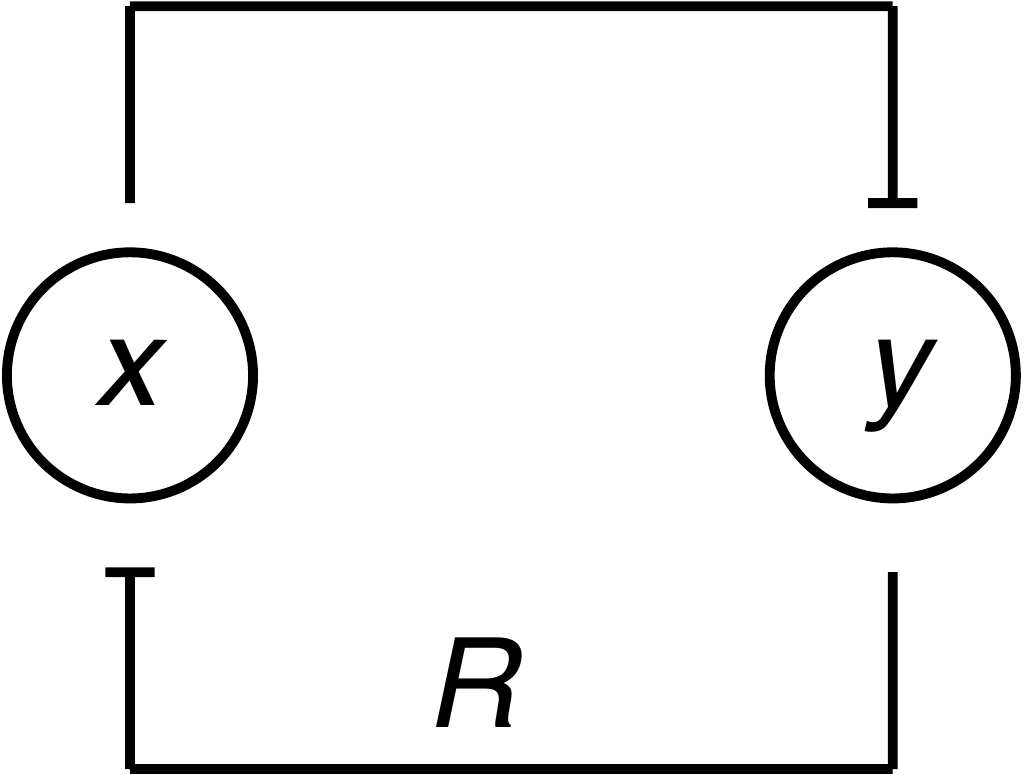}
\caption{The toggle switch. The expression level of gene $x$  is downregulated by gene $y$ upon binding with the repressor R. We make the assumption that when active, both genes synthesize the proteins $x$ and $y$ in one single step of combined transcription and translation.}
\end{figure}
 
As an application of our methodology we consider the toggle switch, a  {positive} feedback loop of two genes mutually repressing each other (Fig. 1) \cite{Gardner00}. The dynamic equations describing each of the synthesized protein have the form (\ref{Gtogsw}), namely:
\BEA
\label{togswa}
\frac{dx}{dt} &=&g(y,R)-k_1x=\frac{g_1} {1 + \rho_1  {(}R y {)}^{\beta_1}}-k_1 x\\
\label{togswb}\frac{dy}{dt} &=&f(x)-k_2 y= \frac{g_2} {1 + \rho_2 x^{\beta_2}}-k_2 y
 \EEA
Here, we have assumed that protein synthesis occurs in one single stage of merged transcription and translation, which, as discussed, is valid when the half-life of mRNA species is much shorter than the half-life of protein species \cite{Rosenfeld02}. Further to standard degradation dynamics for $x$ and $y$ proteins, characterized by degradation rates $k_{1}$ and $k_{2}$ respectively, the structure of the equations is based on two Hill-like functions, which account for each of the two synthesised proteins to act as a repressor on the other gene. Here parameters $\rho_{1,2}$ are the association constants for the binding of $x$ ($y$, respectively) to the promoter of the target gene, $g_{1,2}$ are the maximal expression rates of each gene, and $\beta_{1,2}$ are cooperativity parameters representing homodimerization of the relevant protein.

The bifurcation diagram of this system has been analyzed in detail in \cite{Gardner00}. The system exhibits deterministic multistable behavior for $\beta_1,\beta_2>1$, and it is otherwise monostable. To probe the effect of noise in a situation of deterministic monostability, we therefore set in the following $\beta_1=\beta_2=1$. In this way any transition to bimodal dynamics will only be due to the effect of noise, accounting thereby for the possible emergence of a noise-induced transition in the system.

To explore this further, we have made the assumption that regulation of $x$ happens upon formation of a heterodimer of $y$ with the external (repressive) transcription factor $R$. Hence fluctuations in $R$ will correspond to insertion of nonlinear extrinsic noise in the system, and will be amenable to be studied with the methodology developed here.  

 \subsection{Perturbative calculation}
As for the more general case, let us  add Eq. (\ref{gennetb}) to the  equations (\ref{togswa})  and (\ref{togswb}) for the toggle switch   { and let us  assume  $\tau_x = 1/k_1>  \tau_y=1/k_2$. We then} redefine time as $t \to t/\tau_x$, we obtain:
\BEA
\frac{dx}{dt} &=&\eps \gamma(y,R)-x=\eps \frac{\gamma_1} {1 + \rho_1 R y}- x \ ,\\
 \frac{dy}{dt} &= &\eps f(x)-\kappa y= \eps\frac{\gamma_2} {1 + \rho_2 x}- \kappa y \ ,\\
 \frac{d R}{dt}&=&\eps\mu(R)+\sqrt{ \eps D } \nu(R) \xi(t) \ ,
\EEA
where  $\eps=\tau_x/\tau$ , $\gamma_{1,2}=g_{1,2} \tau$  and  $\kappa=k_2/k_1\simeq 1$. For simplicity we here assume $\kappa=1$.

We also assume that the timescale $\tau$ of the dynamics of $R$ satisfies $\tau \gg \tau_x ,\tau_y $  and therefore $\eps$ is a small parameter. We are in this way assuming that the dynamics of $R$ is slower than that of $x$ and $y$.

With $L_{\rm FP}$ defined as in Eq. (\ref{FP}), the Fokker-Planck (\ref{totalb0}) becomes
\begin{eqnarray}
\label{totalb}
\nonumber \frac{\partial}{\partial t} W_t(x,y,R)&=&\left[-\frac{\partial}{\partial x} \left( \eps \gamma(y,R) - x\right)-  \frac{\partial}{\partial y} \left( \eps f(x) - y\right) \right.\\
&&\left. \vphantom{  -\frac{\partial}{\partial x} } +\eps L_{\rm FP}(R) \right] W_t(x,y,R) \ ,
\end{eqnarray}
whose stationary solution satisfies 
\BEA
\label{sstate}
 \nonumber &&\left[\frac{\partial}{\partial x} \left(x- \eps \gamma(y,R) \right)+ \frac{\partial}{\partial y} \left( y-\eps f(x) \right)
 + \right.\\
 &&  \left. \vphantom{\frac{\partial}{\partial x} }+\eps L_{\rm FP}(R) \right]W(x,y,R)=0
 \EEA

\subsection{Zeroth order solution}
Assuming the perturbative expansion (\ref{expansionZ}), Eq. (\ref{nabla}) becomes 
\BEQ
\left(\frac{\partial}{\partial x}x+\frac{\partial}{\partial y} y\right) W^{(0)}(x,y,R) =0 \ ,
\EEQ
which immediately leads to the zeroth order solution:
\BEQ
W^{(0)} (x,y,R)=\delta(x)\delta(y)\bar{W}(R) \ .
\EEQ

\subsection{First order correction}
We now proceed with the calculation of first order corrections $W^{(1)}(x,y,R)$. Upon substitution of (\ref{expansionZ}) in (\ref{sstate}) and by collecting terms up to first order in $\eps$, we obtain 
\BEA
-\frac{\partial}{\partial x} \left( \gamma(y,R)W^{(0)} - x W^{(1)}\right)&-&\frac{\partial}{\partial y} \left( f(x)W^{(0)} - y W^{(1)}\right) \nonumber \\
&+& L_{\rm FP}(R) W^{(0)}=0
\EEA
which leads to the solution
\BEA
&&\hspace{-0.2cm}W^{(1)}(x,y,R) \nonumber \\
&&\hspace{0.1cm}= - \bar{W}(R) \left( \vphantom{\sum}\delta'(x) \delta(y)\gamma(y,R)+\delta'(y)\delta(x)f(x)\right), \label{W1}
\EEA
as could be derived directly from Eq.(\ref{W1z}).
Following our general procedure, we can write to the first order in $\eps$:
\BEA
\label{firstSol}
W(x,y,R) &=& W^{(0)}+\eps W^{(1)} \nonumber \\
&=&\bar{W}(R) \left[\left(  \delta(x)-\eps \gamma(y,R) \delta'(x) \right) \delta(y) \right.\nonumber \\
&&\qquad \qquad \quad \; \; - \left. \delta'(y) \delta(x) f(x)  \right] \nonumber \\
&=& \bar{W}(R) \delta(x-\eps \gamma(y,R))\delta(y-\eps f(x)) \ ,
\EEA
analogous to Eq. (\ref{firstSolz}). In fact, if we let 
\BEQ
\tilde{\delta}(x,y)=\left(\vphantom{\sum}\delta(x)-\eps \gamma(y,R) \delta'(x) \right) \delta(y) - \eps \delta'(y) \delta(x) f(x) \ , 
\EEQ
then from the distributional point of view 
\BEA
&&\hspace{-0.4cm}\iint  dxdy \tilde{\delta}(x,y) h(x,y) \nonumber \\
&&\hspace{0.2cm}=h(0,0)+\eps \left(\gamma(y,R) \frac{\partial}{\partial x}h(0,0) + f(x)\frac{\partial}{\partial y}h(0,0) \right)\nonumber \\
&&\hspace{0.2cm}=h(\eps \gamma(y,R),\eps f(x))+O(\eps^2)
\EEA
and since 
\BEA
&&\hspace{-0.2cm}h(\eps \gamma(y,R),\eps f(x)) \nonumber \\
&&\hspace{0.3cm}=\iint dx dy \delta(x-\eps \gamma(y,R))\delta(y-\eps f(x))h(x,y)
\EEA
 for any $h(x,y)$, the last equality in (\ref{firstSol}) follows.
This result coincides at this order with the  {noise filtering approach} proposed in \cite{Ochab-Marcinek10} with $p(x)$ stemming from $\bar{W}(R)$ via 
the change of variables $R \to x$ induced by the (analytically solvable in this case) condition $x-\eps \gamma(\eps f(x),R)=0$

\subsection{Second order correction}
We can now carry out the evaluation of the second order correction by computing $L_{\rm FP} (R) W^{(1)}$. Eq. (\ref{L1z}) becomes
\BEA
\label{L1}
\nonumber L_{\rm FP}(R) W^{(1)}(x,y,R) &=&-\delta'(x)\delta(y) L_{\rm FP}(R) \left(\gamma(y,R) \bar{W}(R)\right)\\
 &=&-\delta'(x)\delta(y) G(y,R) \ ,
\EEA
where
\BEA
G(y,R)& =&L_{\rm FP}(R) \left(\gamma(y,R) \bar{W}(R)\right)  \\
&=& - (\mu(R)-3 D\nu(R) \nu'(R))\bar{W}(R)\partial_R\gamma(y,R)+ \nonumber \\
 &+&D \nu(R)^2 (2\bar{W}'(R)  \partial_R \gamma(y,R) +\bar{W}(R) \partial_R^2\gamma(y,R) ).  \nonumber
\EEA
Eq. (\ref{new}) now becomes
\BEA
\nonumber -\frac{\partial}{\partial x} \left( \gamma(y,R)W^{(1)} - \frac{x}{2} W^{(2)}\right)&-&\frac{\partial}{\partial y} \left( f(x)W^{(1)} - \frac{y}{2} W^{(2)}\right)\\
&+& L_{\rm FP}(R) W^{(1)}=0 \ ,
\EEA
which, using (\ref{L1}),  can be rewritten as
\BEA
\label{secord}
\nonumber \hspace{-0.5cm}-\frac{\partial}{\partial x} \left( \gamma(y,R)W^{(1)} - \frac{x}{2} W^{(2)}\right)&-&\frac{\partial}{\partial y} \left( f(x)W^{(1)} - \frac{y}{2} W^{(2)}\right)\\
&+&\delta'(x)\delta(y) G(y,R)=0 \ .
\EEA
The form of the solution $W^{(2)}(x,y,R)$ is analogous to (\ref{secordbz}), namely given by
\BEA
\label{secordb}
\nonumber W^{(2)}(x,y,R) &=& \bar{W}(R)\left(A_{10} \delta'(x)\delta(y)+A_{11}\delta'(x)\delta'(y) \right.\\
&+& \left .A_{20}\delta^{\prime \prime}(x)\delta(y)+A_{02}\delta(x)\delta^{\prime \prime}(y)\right)
\EEA
Inserting this expression in (\ref{secord})  one  obtains the following conditions on the coefficients $A_{ij}$ 
\BEA
\label{secordc}
A_{10}&=&-2\bar{G}(y,R)\ ,\;\;\; A_{20}=\left(\gamma(y,R)\right)^2 \ ,\nonumber \\
\nonumber  A_{11}&=&2\gamma(y,R)f(x)\ ,\;\;\;  A_{02}=\left(f(x)\right)^2 \ .
\EEA
with $\bar{G}=G/\bar{W}$, as could be derived as well directly from Eqs. (\ref{secordcz2}) with the condition ${\bf \tilde{A}}=0$.
The ansatz for the steady state solution  for the probability density $W(x,y,R)$ that coincides up to the second order in $\eps$  with the perturbative evaluation shown above   is therefore
\BEA
\label{ansatz}
\nonumber W(x,y,R)&&=\bar{W}(R)\delta\left(x-\eps \gamma(y,R)-\eps^2 \bar{G}(y,R)\right) \\
&&\times \delta\left(y-\eps f(x)\right) +O(\eps^3).
\EEA

\subsection{Log-normal Noise}
The power of the approach that we propose here relies also on the fact that we can assume different types of noise, as long as these can be represented in terms of the Langevin dynamics (\ref{gennetb}). By following \cite{Aquino20}, we focus on log-normal noise, which in contrast to Gaussian noise preserves the positivity of parameter values on which it acts. It also describes well extrinsic fluctuations \cite{Rosenfeld05,Shahrezaei08} and universal behaviours in bacteria and yeast \cite{Salman12,Brenner15}.

The stationary log-normal distribution
\BEQ
\bar{w}(R)=\frac{1}{R\sqrt{2\pi}} \exp{\left[-\frac{1}{2D}   \left( \log\frac{R}{\bar{R}}  +\frac{D}{2}\right)^2 \right]} \label{lognormdistr}
\EEQ
can be derived \cite{Aquino20} by defining the stochastic process  
\BEQ
\label{outolognorm}
R(t)=\bar{R}e^{\eta(t) } e^{-D/2} ,
\EEQ
where  {$\eta(t)$ is the standard Ornstein-Uhlenbeck noise,
\BEQ
\frac{d\eta}{dt} = -\frac{\eta}{\tau} + \sqrt{\frac{D}{\tau}} \xi(t),
\EEQ  
with $\xi(t)$ zero average Gaussian white noise, and $\bar{R}=\langle R(t) \rangle$, since $\langle e^{\eta(t)} \rangle = e^{D/2}$. Hence, the stochastic process (\ref{outolognorm}) obeys 
\BEQ
\label{lognor2}
  \frac{d R}{dt}=-\frac{R}{\tau} \log {\left(\frac{R}{\bar{R}}  +\frac{D}{2}\right)} +\sqrt{ \frac{ D}{\tau}} R \xi(t) , 
  \EEQ
which corresponds to (\ref{gennetb}) with the choice $\mu(R) = -R \log (R/\bar{R}  + D/2)$ and $\nu(R) = R$. Eq. (\ref{generalpf}) allows then the derivation of (\ref{lognormdistr}).}

\subsection{Comparison with numerical simulations}
 
The joint probability density $p(x,y)$ for the protein distributions can be obtained by marginalizing (\ref{ansatz}) with respect to $ R$:
\BEQ
\label{pxy}
p(x,y)=\int dR W(x,y,R) \ .
\EEQ
The probability density $p_x(x)$ for the  distribution  of protein $x$ can  then be obtained by marginalizing (\ref{pxy}) with respect to $y$. 
In this case it is simpler to first integrate in $y$ after replacing (\ref{ansatz}) in (\ref{pxy}), we obtain
\BEA
\label{ansatz2s}
\nn p_x(x)&=&\int dR \bar{W}(R)\delta\left(x-\eps \gamma(\eps f(x),R)-\eps^2 \bar{G}(\eps f(x) ,R)\right)\\
 &=&\int dR \bar{W}(R)\delta(z(x,R))
\EEA
and further integration with respect to $R$ amounts to a change of variable $R=r(x)$
where the function $r(x)$ is defined implicitly by the equation
\BEQ
z(x,R)=x-\eps \gamma(\eps f(x),R)-\eps^2 \bar{G}(\eps f(x) ,R)=0 \ .\label{z1}
\EEQ

We notice that in this particular case we can also develop easily the third order correction. This is obtained following the same procedure used for the second order (see Appendix for details) leading to the following  modified expression for the
steady state solution for the probability density $W(x,y,R)$: 
\BEA
\label{ansatz2b}
\nonumber &&W(x,y,R)=\bar{W}(R)\delta\left(x-\eps \gamma(y,R)-\eps^2 \bar{G}(y,R)+ \right. \\
&&\left. -\eps^3 \bar{G_3}(x,y,R)\right)\delta\left(y-\eps f(x)\right) +O(\eps^4) \ .
\EEA

\begin{figure}[t!]
\includegraphics[height=5.8 cm,width=7.8 cm, angle=0]{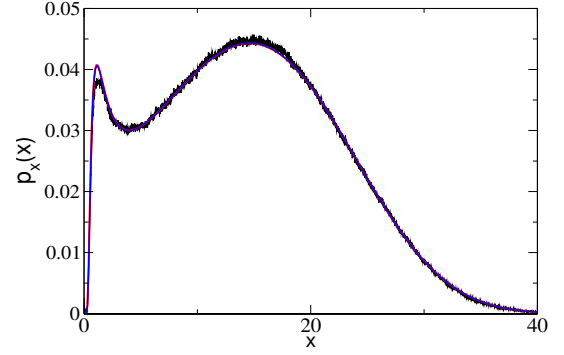}
\hspace*{-0.7cm}\includegraphics[height=5.8 cm,width=8.6 cm, angle=0]{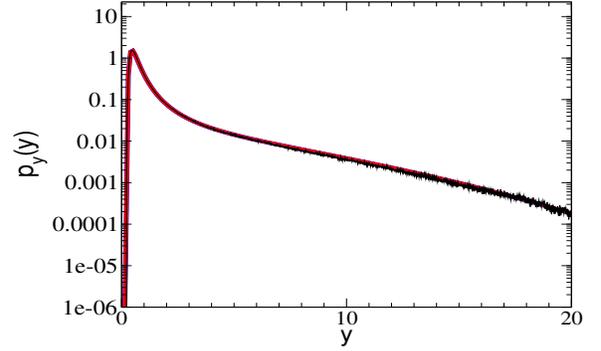}
\caption{Probability distributions $p_x(x)$ (top  {panel}) and $p_y(y)$ (bottom  {panel})
as given  by numerical simulation (black line) compared to theoretical solution (\ref{solana}) for $p_x$ and (\ref{py}) for $p_y$, respectively at the first order (red line), second order via Eq. (\ref{z1}) (blue line), and third order via Eq. (\ref{z2}) (blue continuous line).These solutions are indistinguishable.  {Same colours and line styles  are used in both panels}.   Here $\alpha=10, \bar{R}=0.87$nM, $g_0=0.01$, $\tau=10^7$, $\eps=0.001$, $D=0.012$. } \label{standard}
\end{figure}Here, we have used (see Appendix):
\BEQ
 \bar{G_3}(x,y,R)= \bar{G}_M(y,R)+\frac{\bar{G}_2(y,R)}{ 2 x} \ ,
 \EEQ
and therefore by integrating over $y$ we obtain
 \BEQ
\label{ansatz3s}
p_x(x)=\int dR \bar{W}(R)\delta(z(x,R))
\EEQ
with
 \BEA
 \label{z2}
    z(x,R)&=& x-\eps \gamma(\eps f(x),R)-\eps^2 \bar{G}(\eps f(x) ,R)+\\
 \nonumber &-&\eps^3\bar{G_3}(x,\eps f(x),R) ,  
 \EEA 
which replaces (\ref{z1}).

According to the implicit functions theorem, from the condition $z(x,R)=0$, it follows that a function $r(x)
=R$ exists such that 
\BEQ
\label{solana}
p_x(x)=\bar{W}(r(x))\left|\frac{dr}{dx}\right| \ , 
\EEQ
where $dr/dx$ can be computed by applying the implicit functions theorem    either to  (\ref{z1}) for the calculation up to the second order, or to (\ref{z2}) for the third order. Hence: 
\BEQ
\frac{dr}{dx}=-\frac{\partial_x z(x,R)}{\partial_R z(x,R)} \ ,
\EEQ
which allows us to evaluate explicitly $p_x(x)$  for the toggle switch.

\begin{figure}[t!]
\includegraphics[height=5.8 cm,width=7.8 cm, angle=0]{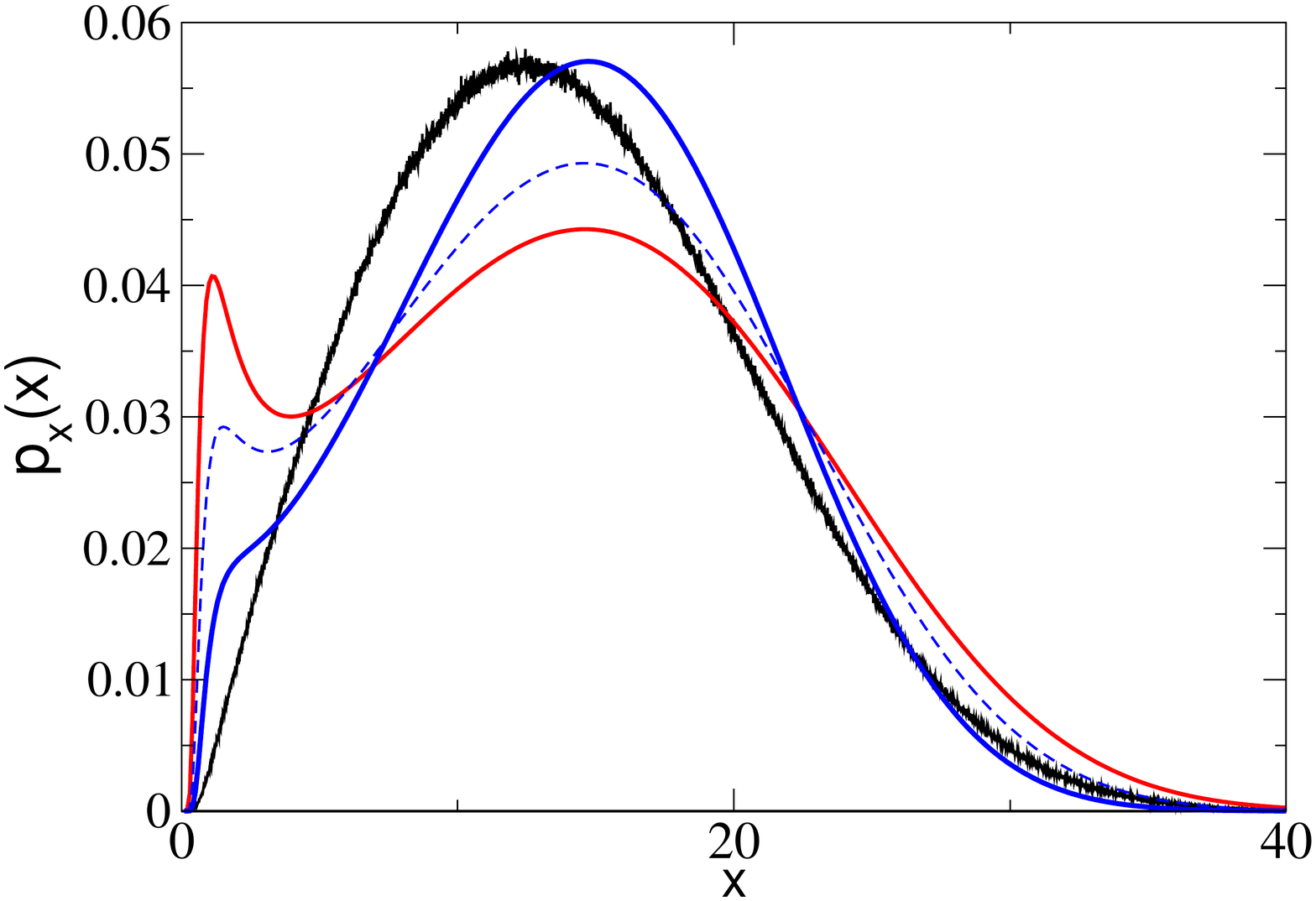}
\hspace*{-0.7cm}\includegraphics[height=5.8 cm,width=8.6 cm, angle=0]{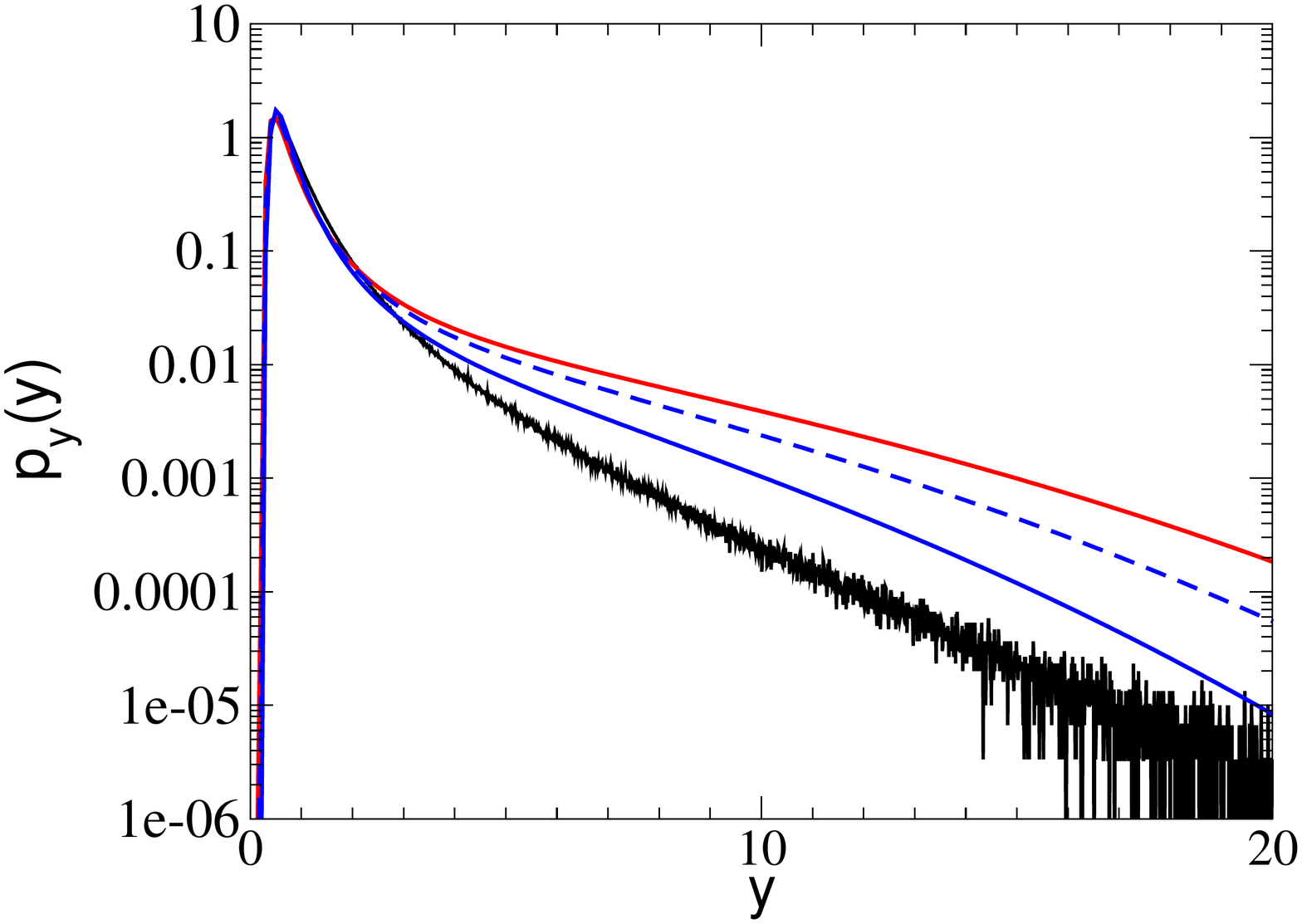}
\caption{Probability distributions $p_x(x)$(top  {panel}) and $p_y(y)$ (bottom  {panel})
as given  by numerical simulation (black line) compared to theoretical solution (\ref{solana}) for $p_x$ and (\ref{py}) for $p_y$, respectively at the first order (red lines), second order via Eq. (\ref{z1}) (blue dashed lines), and third order via Eq. (\ref{z2}) (blue continuous lines).  {Same colours and line styles  are used in both panels}.  Including the third order correctly shows only one mode for $p_x(x)$ (blue continuous line, top panel). Here $\alpha=10$, $\bar{R}=0.87$nM, $g_0=0.01$, $\tau=10^5$, $\eps=0.1$, $D=0.012$. } \label{nonstandard}
\end{figure}

Finally, in order to derive
\BEQ
p_y(y)=\int dx \int dR W(x,y,R) \ ,
\EEQ
it suffices to notice that at steady state we have exactly $y=\eps f(x)$, and therefore $p_y(y)$ in this case can be derived  directly from $p_x(x)$ via a change of variable as 
\BEA
\label{ansatz2}
p_y(y)&=&\frac{1}{\eps}p_x\left(f^{-1}\left(\frac{y}{\eps}\right)\right)\left| \frac{df^{-1}}{dy} \right| \nonumber \\
&=&\frac{p_x(f^{-1}(y/\eps))}{\eps \left| f'(x=f^{-1}(y/\eps)\right|} \ . \label{py}
\EEA

Equations (\ref{solana}) and (\ref{py}) so constructed can then be compared at any given perturbative order with direct stochastic simulations, in order to highlight the predictive power of the theoretical approach here developed.

We perform direct stochastic simulations of fluctuatons in $R$ by using the standard approach to generate an Ornstein-Uhlenbeck process, converting it via Eq. (\ref{outolognorm})  into log-normal noise and then by performing ordinary integration of the equations (\ref{togswa}) and (\ref{togswb}). The probability densities are obtained by averaging a single trajectory over long times. Parameters are set so as to reproduce biologically relevant conditions. In all simulations and in the analytical solution, we fix the maximal gene expression rates at the values  {$g_1 = g_2 = g_0 = 10^{-2} s^{-1}$}, which is representative of typical gene expression in bacteria \cite{Thattai01}, and the degradation rates $k_1 = k_2 = 10^{-4} s^{-1}$, which are representative of proteins' half-life in the range of 2 hours. We also fix the dissociation constants of transcription factors to DNA  {$\rho_1 = 10 nM^{-1}$ and $\rho_2 = 1 nM^{-1}$, and $\bar{R} = 0.87 nM$} \cite{Aquino20}. 

We then focus on two different values of $\eps$ in order to explore the contribution of different perturbative orders. Since we assume $k = 10^{-4} s^{-1}$, choosing $\tau = 10^{7} s$ corresponds to assume $\eps=0.001$, while the choice $\tau = 10^{5} s$ will reflect a larger value of $\eps$, namely $\eps=0.1$. It is this second regime, when the timescale separation between dynamics of $x$ and of the $R$ fluctuations is less pronounced, that we predict that higher order terms in the perturbative expansion will become relevant. 

In Fig. \ref{standard} we show the comparison between perturbative expansion and stochastic simulations for $\eps=0.001$. The theoretical analysis predicts bimodality for the concentration $x$, and this is confirmed excellently by the numerical result. In this case  first order  solution Eq. (\ref{firstSol}), second order solution via Eq. (\ref{z1}), and third order solution via Eq. (\ref{z2}) practically coincide.

In Fig. \ref{nonstandard}  (top panel) we show a particular case
where using the first order solution Eq. (\ref{firstSol}) leads erroneously to conclude that bimodality emerges, while including second and third order correction gives a quantitatively and qualitatively correct description of the properties of the density function, correctly  predicting a unimodal distribution for both $x$ and $y$ protein concentrations.

\subsection{Noise-induced transitions: Partial bimodality}

\begin{figure}[t!]
\centering
   \includegraphics[width=0.43\textwidth]{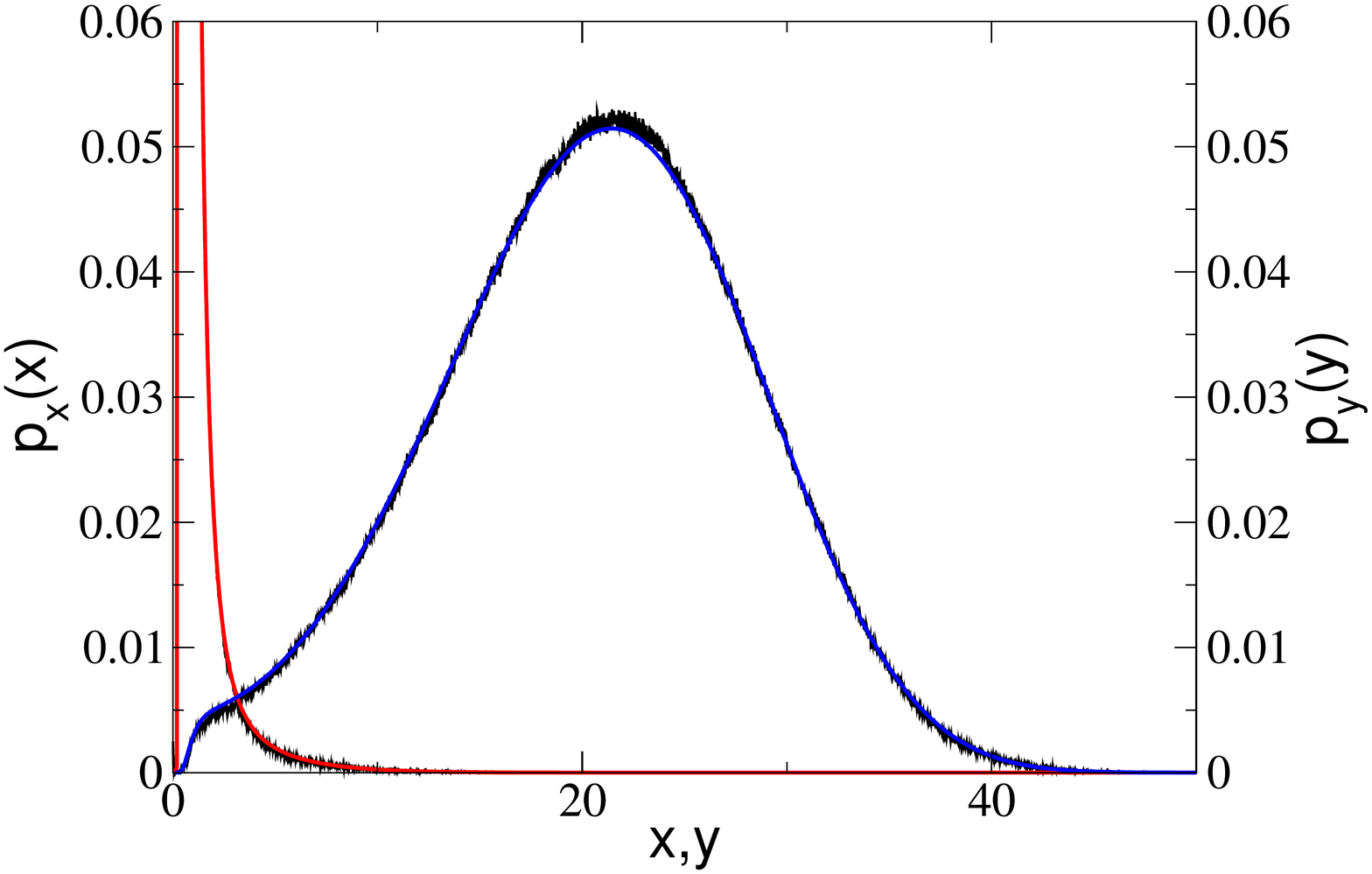}
   \includegraphics[width=0.43\textwidth]{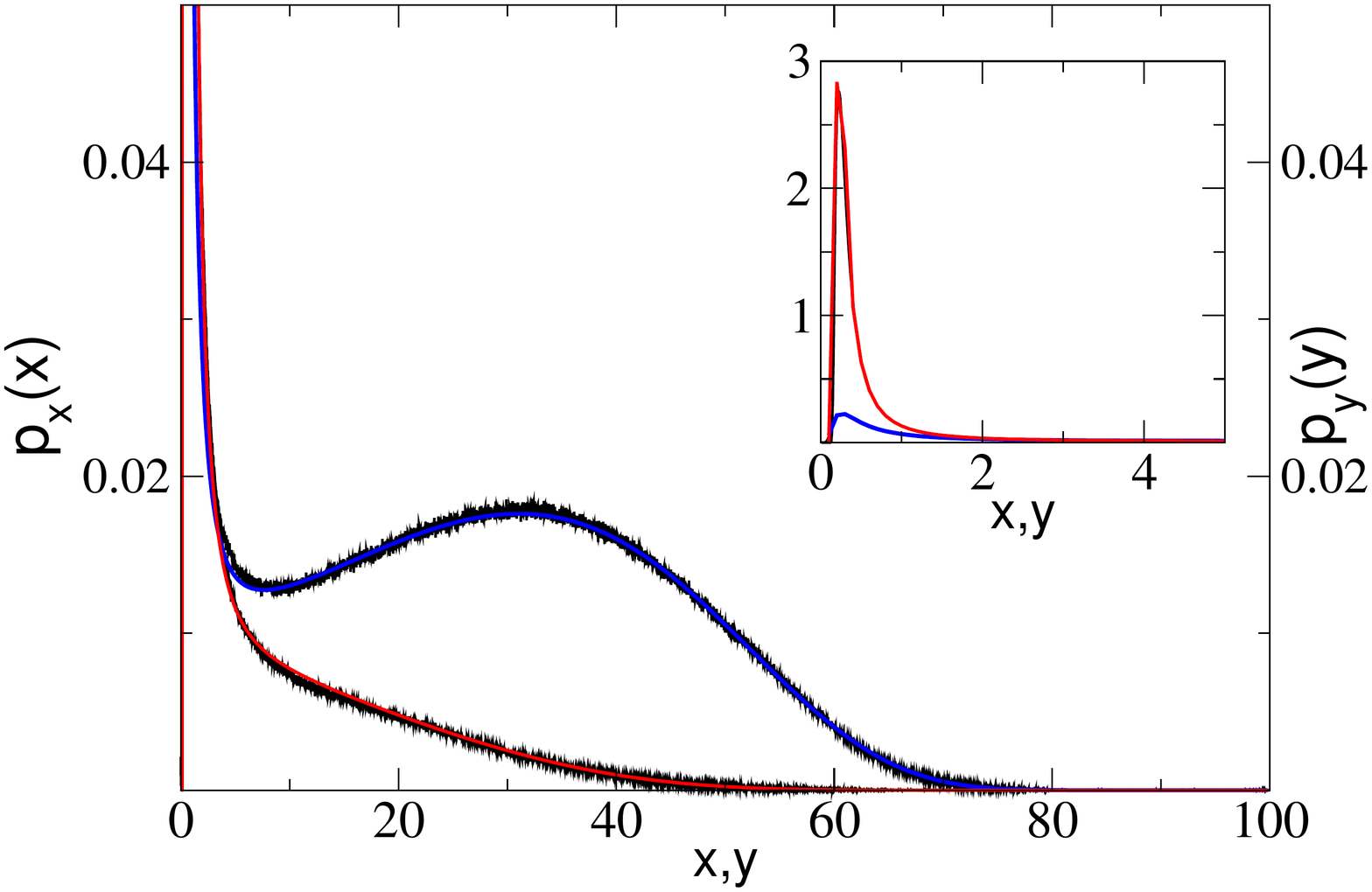}
      \includegraphics[width=0.43\textwidth]{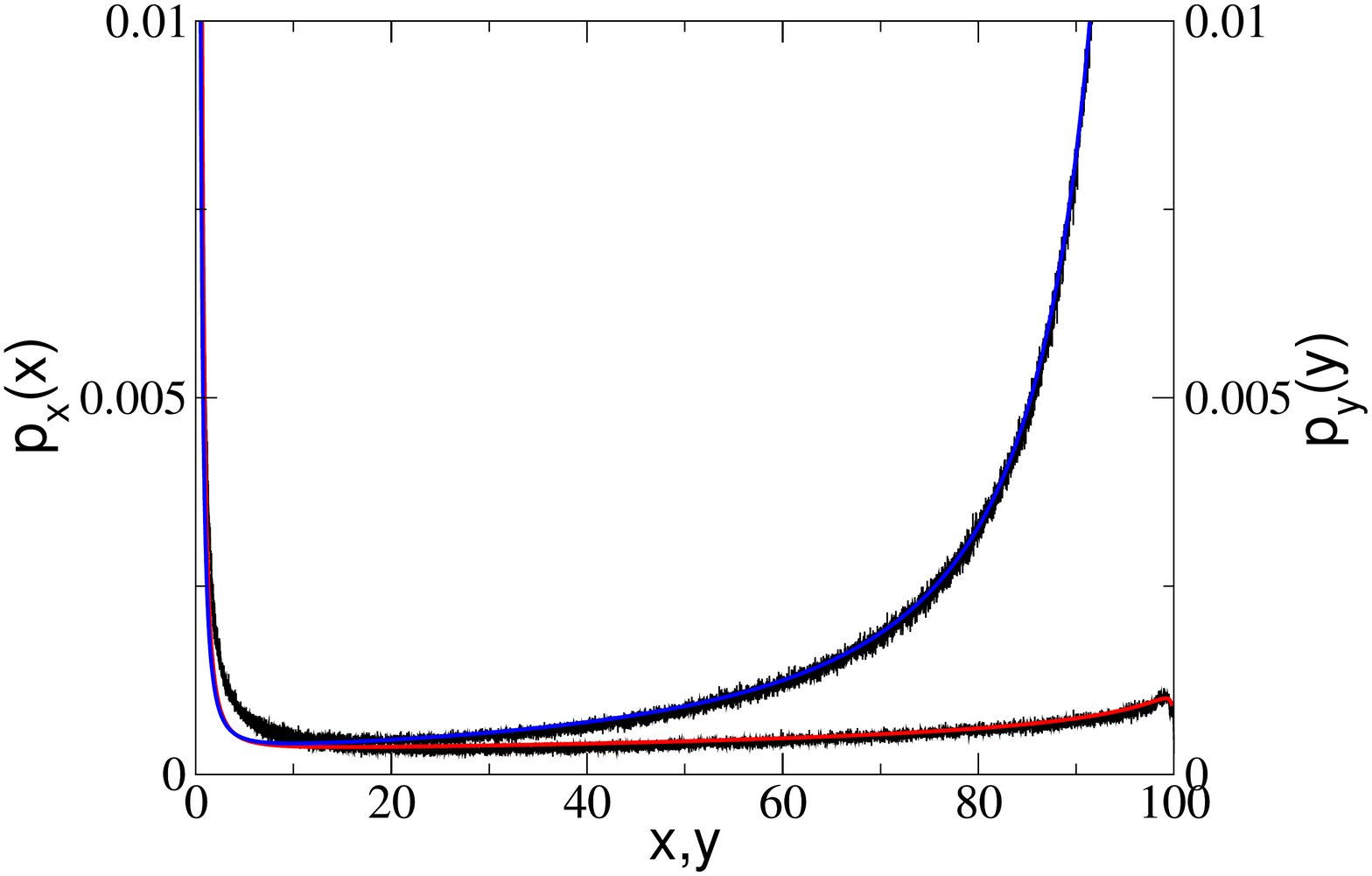}
\caption{Noise-induced transitions and partial bimodality in the toggle switch. The distributions $p_x(x)$ and $p_y(y)$ as given  by numerical simulations (black lines) compared to  analytical solutions  (\ref{firstSol}) (red line and blue line for $p_y(y)$ and $p_x(x)$ respectively). Here $\alpha=10, \bar{R}=0.8nM, g_0=0.01, \tau=10^7, \eps=0.001$, from top to bottom $D=0.01$,  {0.1}, $10$.  } \label{partialbim}
\end{figure}

The analysis and simulations shown in Fig. \ref{standard} highlight the emergence of a particular phenomenon, which we name 'partial bimodality', whereby bimodal behaviour emerges for the protein concentration $x$, but not for $y$. This is clearly unexpected in standard toggle switches affected by intrinsic noise only, as in this case the bimodality would instead affect simultaneosly both genes involved.

In order to obtain further insight into these dynamics, and clarify the nature of the noise-induced transition here identified, we perform simulations and solve the corresponding first order dynamics, for three different values of noise intensities.

In Fig. \ref{partialbim} we show simulations and analytical predictions for the toggle switch in a regime where at low noise intensity ($D=0.01$) the dynamics are unimodal. In this case we observe that increasing the noise intensity up to $D=10$ bimodal behaviour emerges for both genes. However, interestingly we also observe that for intermediate values of noise,  {$D\simeq 0.1$}, a regime exists where bimodality emerges only for one gene and not the other. 

\begin{figure}[t!]
\includegraphics[width=0.43\textwidth]{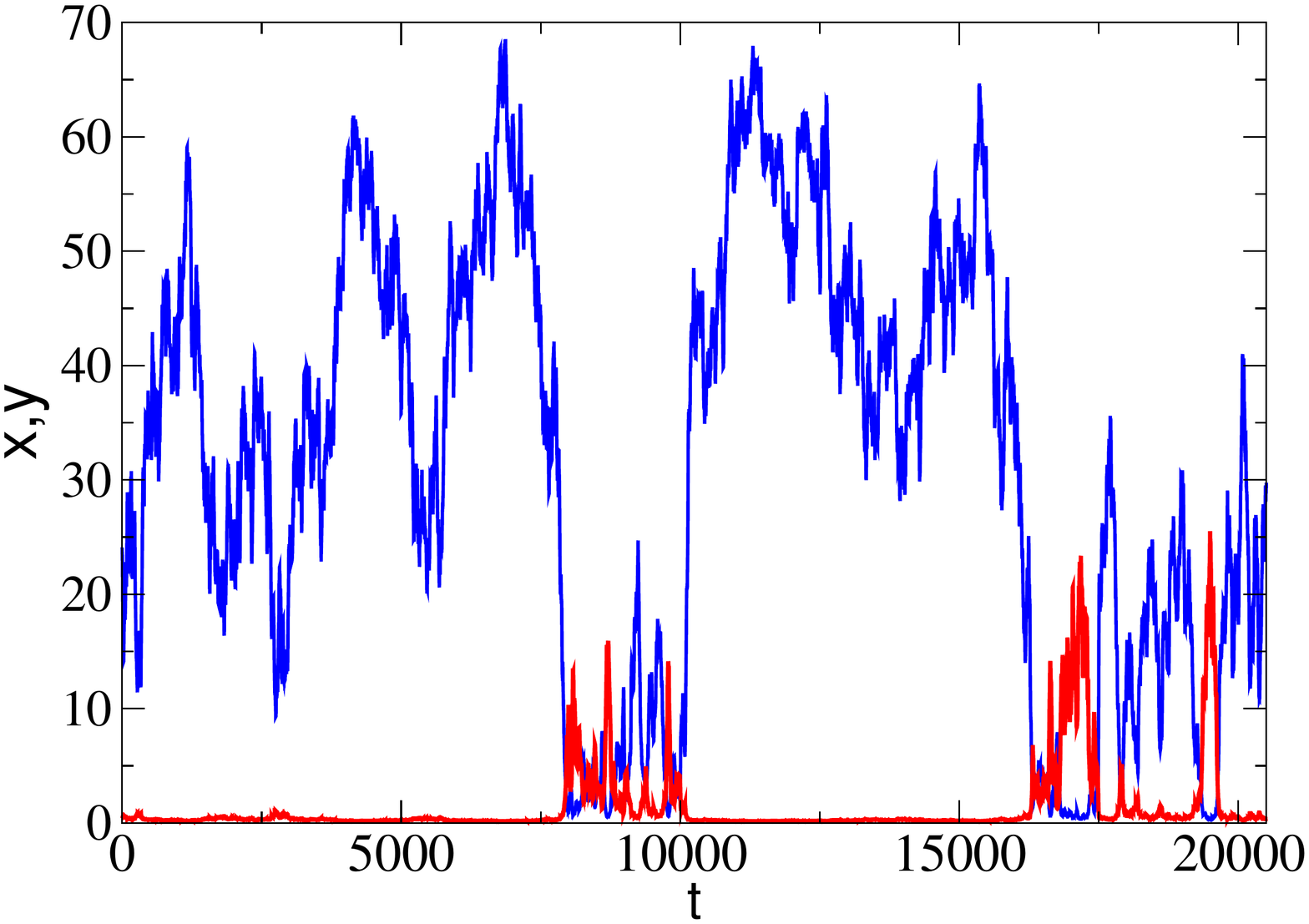}
\includegraphics[width=0.43\textwidth]{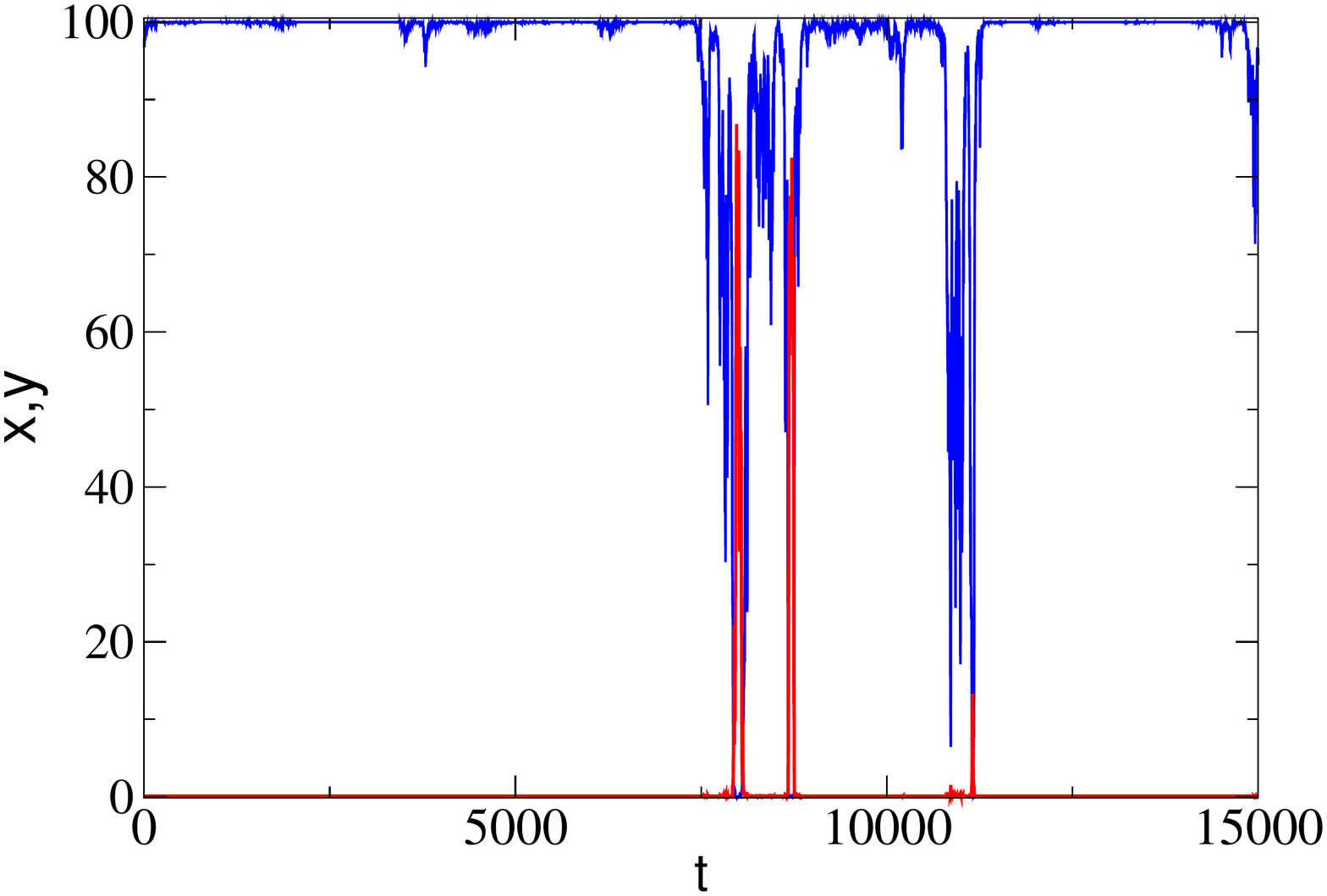}
\caption{ {Exemplary trajectories for the variable $x$ (blue line) and $y$ (red line) for parameters $\alpha=10$, $\bar{R}=0.8$nM, $g_0=0.01$, $\tau=10^7$, $\eps=0.001$ and  $D=0.1$ (upper panel, corresponding to distributions with partial bimodality in middle panel in Fig. \ref{partialbim})  and $D=10$ (lower panel, corresponding to bimodal distributions in bottom panel in Fig. \ref{partialbim}).}}
\label{trajec}
\end{figure}

 {For this case we show two exemplary trajectories for variables $x$ and $y$ in the upper panel of Fig. \ref{trajec}. While the variable  $x$ makes complete transitions between  a high value  around the second mode ($x = 32.5$)  and low values close to the first mode ($x=0.24)$, the variable $y$ remains more consistently around $y=0.2$, as testified by the higher peak of its distribution at the origin as compared to that of $p_x$ (see inset in middle Figure \ref{partialbim}). In correspondence of transitions of the $x$ variable to the low state, excursions to slightly larger values of $y$ are permitted, but their amplitude is not large enough to exit the basin of attraction of the low $y$ state. In the bottom panel of Fig. \ref{trajec} instead, the case where bimodality occurs for both species is shown.  These traejctories correspond to the bimodal distributions in the bottom panel of Fig. \ref{partialbim} with modes in $x,y=0$  and in $x= 100, y=99$.  Since  $p_x$ has a higher peak in $x=100$  and $p_y$  has a higher peak in  $y=0$  these species spend most of their time around these values with sporadic transitions to values around $x=0$ and $y= 99$ respectively. The amplitude of the fluctuations of the $y$ variable allows in this case the variable $y$ to exit the basin of attraction of its high state.}

 {We attribute the different noise amplitudes experienced by the variables $x$ and $y$ to a noise propagation effect, whereby the fluctuations of the external parameter $R$ are perceived differently by the two genes. In fact, the variance of the distribution of $x$ is consistently larger of the variance of the distribution of $y$, for all input noises on $R$ (parameterized by $D$), as shown in Table 1:}

\begin{table}[ht]
\centering 
\begin{tabular}{c c c c} 
\hline
$D$ & $0.01$ & $0.1$ & $10$\\ [0.5ex] 
\hline 
$\sigma_{x}$ & 56.62 & 325.31 & 590.67\\ 
$\sigma_{y}$ & 0.88 & 58.57 & 191.17\\ [1ex] 
\hline 
\end{tabular}
\caption{Variances of $x$ and $y$ distributions corresponding to the probability distributions shown in Fig. \ref{partialbim}. Parameter values are set as in Fig. \ref{partialbim}.} 
\end{table}

 {While $x$ is under the direct influence of fluctuations of $R$, gene $y$ feels those fluctuations, propagated through the repressive action from $x$, at a lower level. This means that while gene $x$ can be above threshold for transitions to bimodal behavior, gene $y$ may still remain below the threshold, and therefore may still exhibit unimodal behaviour. This mechanism is in agreement with the result of some extensive studies performed in \cite{Hornung08}, which demonstrated that positive feedback loops limit the range of fluctuations of the target gene.}

\section{Conclusions}

In this paper we have introduced an efficient methodology to deal with nonlinear extrinsic noise when the correlation time of the fluctuations is large. The derivation here proposed is based on a timescale expansion of the master equation and reproduces to the first order the phenomenological approach based on  { the nonlinear noise filtering}  introduced in \cite{Ochab-Marcinek10}.  Importantly, our perturbative approach allows one to go beyond  the dynamical derivation presented in \cite{Aquino20} and obtain correct predictions even in a regime when the timescales characterizing fluctuations and gene dynamics are comparable, with the lowest perturbative orders failing to capture the qualitative dynamical features of the system.

We also find that the toggle switch is capable of a partial bimodality, in that while one of the two participating genes shows full bimodal behaviour, the other gene still preserves unimodality. This feature is intermediate in terms of noise intensities to the two extreme cases of no bimodality for any of the two genes, to full bimodality for both of them. As a follow-up to this observation, it will be interesting to analyze the effect of simultaneous intrinsic and extrinsic noise on the system. We expect that the interplay between intrinsic and extrinsic noise would provoke even richer dynamics, in which the balance between noise intensities will be crucial in determining the structure of the phase space.

The novel mechanism for noise-induced transitions here described  and the methodology introduced are promising and worthy of further analysis and experimental validation. They will further our understanding of the fundamental and active role played by noise in biological systems.

\section*{ACKNOWLEDGMENTS}
This study has been funded by the UK Biotechnology and Biological Sciences Research Council (BBSRC), under grant no. BB/L007789/1.

For the purpose of open access, the authors have applied a Creative Commons Attribution (CC BY) licence to any Author Accepted Manuscript version arising.

\section*{DATA AVAILABILITY}
All details of the mathematical models developed and the data deriving from their simulation are available in the article. 

\section*{APPENDIX: EVALUATION OF THIRD ORDER PERTURBATIVE CORRECTION FOR THE TOGGLE SWITCH}

We can evaluate the third order correction in the perturbative expansion just by carrying on the procedure introduced in the main text 
to include the third order. We start with $W(x,R)=W^{(0)}+\eps W^{(1)}+\eps^2W^{(2)}/2+\eps^3W^{(3)}/6 $ and this  leads to 
\BEA
\label{order3}
&&-\frac{\partial}{\partial x} \left( \gamma(y,R) W^{(2)}- \frac{x W^{(3)}}{3} \right) \nonumber \\
&&-\frac{\partial}{\partial y} \left( f(x)W^{(2)}- \frac{y W^{(3)}}{3} \right)+L_{\rm FP}(R)W^{(2)}=0.
\EEA
where we know $W^{(2)}(x,y,R)$ from previous calculations. Therefore we can evaluate  $L_{\rm FP}(R) W^{(2)}$, and we obtain
\BEA
&&L_{\rm FP}(R) W^{(2)}(x,y,R)=G_{2}(y,R)\delta''(x)\delta(y) 
\nonumber \\
&&+2f(x)G(y,R) \delta'(x)\delta'(y)-2G_M(y,R)\delta'(x)\delta(y) 
\EEA
with
\BEA
&&\hspace{-0.5cm}G_{2}(y,R)=L_{\rm FP}\left(\gamma^2(y,R)\bar{W}(R)\right)= \nonumber \\
&&\hspace{-0.5cm} \qquad 2(3 D\nu(R) \nu'(R)-\mu(R)) \bar{W}(R)\gamma(y,R)\partial_R\gamma(y,R) \nonumber \\
&&\hspace{-0.5cm}\qquad +D \nu(R)^2 \left[4\bar{W}'(R) \gamma(y,R)  \partial_R \gamma(y,R) \right. \nonumber \\
&&\hspace{-0.5cm}\qquad \left. + 2 \bar{W}(R) \left(\gamma(y,R)\partial_R^2\gamma(y,R) +(\partial_R\gamma(y,R) )^2 \right)\right] \ ,
\EEA
where it can be shown that 
\BEA
&&\hspace{-0.5cm}G_M(y,R)=L_{\rm FP}(R)\left(G(y,R)\right)= \nonumber \\
&&\hspace{-0.2cm}\alpha(R)G(y,R) + \sigma(R)\partial_R G(y,R)+\gamma(R)\partial_R^2 G(y,R) \ ,
\EEA
with
\BEA
\alpha(R)&=&-\left(\mu'(R)-D\nu'(R)^2-D\nu(R)\nu^{\prime \prime}(R) \right) \ ,\\
\sigma(R)&=& -\left.(\mu(R)-3D\nu(R)\nu^{\prime}(R) \right) \ ,\\
\gamma(R)&=& D \nu^2(R) \ .
\EEA
From Eq. (\ref{order3}) we can therefore expect $W_3(x,R)$ of the 
form:
\BEA
&&W_3(x,y,R)=\bar{W}(R) \left(A_{10}\delta^{\prime }(x)\delta(y) +A_{11}\delta'(x) \delta'(y) \right. \nonumber\\
\nonumber && \qquad \left. + A_{12}\delta'(x) \delta^{\prime \prime}(y)+ A_{20}\delta^{\prime \prime}(x)\delta(y)+A_{21} \delta^{\prime \prime}(x) \delta'(y)+\right.\\
&&\qquad \left. +A_{30} \delta'''(x)\delta(y)+A_{03}\delta(x)\delta'''(y)\right) \ . \label{ans3}
\EEA
Replacing Eq. (\ref{ans3}) in (\ref{order3}) and solving for $A_{ij}$ leads to
\BEA
A_{10}&=&-6 \bar{G}_M(y,R)-(3/x)\bar{G}_2(y,R)\\
A_{11}&=&6\bar{G}(y,R) f(x)\\
A_{12}&=&3 \gamma(y,R)f^2(x)\\
A_{20}&=&6\bar{G}(y,R)\gamma(y,R)\\
A_{21}&=-&3\gamma^2(y,R)f(x)\\
A_{30}&=&- \gamma^3(y,R)\\ 
A_{03}&=&- f^3(x) 
\EEA

\end{document}